\documentclass[showpacs,twocolumn,pra]{revtex4}
\usepackage{graphicx}
\usepackage{epsfig}
\usepackage{amssymb}

\def\beq{\begin{equation}}
\def\eeq{\end{equation}}
\def\beqa{\begin{eqnarray}}
\def\eeqa{\end{eqnarray}}

\def\half{\frac{1}{2}}

\def\D{\Delta}
\def\del{\delta}

\def\cH{{\mathcal H}}
\def\cL{{\mathcal L}}

\def\d{\mathrm{d}}

\def\nonum{\nonumber \\}

\def\del{\delta}

\def\nonum{ \nonumber \\}

\newcommand{\ket}[1]{| #1 \rangle}
\newcommand{\bra}[1]{\langle #1 |}

\newcommand{\eqref}[1]{~(\ref{#1})}

\begin{document}

\title{ Relaxation times in an open interacting two-qubit system}

\author{Y. Dubi and M. Di Ventra}

\affiliation{Department of Physics, University of California San
Diego, La Jolla, California 92093-0319, USA} \pacs{...}

\begin{abstract}
In a two-qubit system the coupling with an environment affects
considerably the entanglement dynamics, and usually leads to the
loss of entanglement within a finite time. Since entanglement is a
key feature in the application of such systems to quantum
information processing, it is highly desirable to find a way to
prolonging its lifetime. We present a simple model of an interacting
two-qubit system in the presence of a thermal Markovian environment.
The qubits are modeled as interacting spin-$\half$ particles in a
magnetic field and the environment is limited to inducing single
spin-flip events. A simple scheme allows us to calculate the
relaxation rates for all processes. We show that the relaxation
dynamics of the most entangled state exhibit critical slowing down
as a function of the magnetic field, where the relaxation rate
changes from exponentially small values to finite values in the
zero-temperature limit. We study the effect of temperature and
magnetic field on all the other relaxation rates and find that they
exhibit unusual properties, such as non-monotonic dependence on
temperature and a discontinuity as a function of magnetic field. In
addition, a simple scheme to include non-Markovian effects is
presented and applied to the two-qubit model. We find that the
relaxation rates exhibit a sharp, cusp-like resonant structure as a
function of the environment memory-time, and that for long
memory-times all the different relaxation rates merge into a single
one.
\end{abstract}
\maketitle

 \section{Introduction}
When considering application of quantum information processing, two main ingredients must be considered - entanglement and
decoherence. While entanglement, or non-local coherence, plays a key role in qubit operations \cite{review}, decoherence sets the
limit to which such operations may be performed \cite{Zurek}. Decoherence may result from the interaction of the quantum system
with a dissipative environment \cite{Mintert1}, dramatically affecting the dynamics of the quantum system, and its entanglement
properties. Specifically, it was recently shown \cite{Dodd,YuEberly} that a (non interacting) two-qubit system can be completely
disentangled in a finite time, a phenomenon dubbed "entanglement sudden death". This was followed by a plethora of theoretical
studies of this phenomenon in various situations, most of them treating non-interacting qubits (i.e. a pair of qubits which
interact with each other only in the mediation of the environment), either with a Markovian or non-Markovian environment
\cite{Zagury,Ficek,Lastra,Fanchini,Ikram,Bellomo,Dajka,Cao,Qasimi}. From the experimental side, direct measurements of
entanglement have been performed \cite{experiment1,experiment2}, and the "entanglement sudden death" was observed
\cite{experiment3}.

The loss of entanglement seems to be a generic feature of two-qubit systems \cite{YuEberly}. Our goal is to study a simple system
where entanglement sudden death may be avoided. For this aim we study a simple model of an interacting two-qubit system in the
presence of a thermal dissipative bath. We model the qubits as interacting spin-$\half$ particles in a magnetic field
\cite{Gunlycke} and use the Born-Markov approximation for the system-environment coupling. The environment is assumed to be ohmic
and induce thermal transitions, and it only allows for single spin-flip events (in similarity to spin-boson models \cite{LeHur}).
We calculate analytically the full dynamics of this system, with emphasis given to the different relaxation rates. We point out
here that we use the term "relaxation rates" loosely, to describe the time-scales of both processes which include energy changes
(relaxation) and only coherence loss (decoherence). Both types of processes are inherently present in our calculation scheme. We
show that although the coupling with the environment is characterized by a single relaxation rate, different relaxation rates
emerge for different coherent states. We study the effect of temperature and magnetic field on the different relaxation rates and
find that they may be non-monotonous functions of temperature.

As the main results of this paper, we demonstrate that as a function
of the magnetic field the relaxation rate of the highly-entangled
states abruptly changes from being finite as the temperature
vanishes to being exponentially small. This occurs when one of the
states of the entangled pair is in a meta-stable state, and
indicates that with a proper tuning of parameters the entanglement
may survive very long times even in the presence of a dissipative
environment. We demonstrate the long-lifeness of entanglement by
calculating the concurrence of a specific entangled state, and show
that applying a transverse magnetic field destroys this effect.

Finally, we devise a simple way to account for non-Markovian effects
when calculating relaxation rates. Studying the relaxation rates as
a function of the environment memory-time, we find that some
relaxation rates exhibit a non-monotonic dependence on the
memory-time, with a cusp-like resonance. For long memory-times, the
different relaxation rates merge into a single rate.

\section{Method}

Let us introduce our method for calculating the relaxation times. We
consider a quantum system, characterized by a time-independent
Hamiltonian $\cH$ with $N$ energy levels $E_k,~k=1,2,...N$. The
system dynamics are given by the evolution of the density matrix
$\rho(t)=\sum_{kk'} \rho_{kk'}(t) \ket{k} \bra{k'}$, where $\ket{k}$
are the eigenfunctions of the Hamiltonian.~\cite{Persh} For the above
choice of Hamiltonian, in the Markovian
approximation the evolution of the density matrix is given by a
quantum master equation \cite{VanKampen} ($\hbar=1)$,
 \beq  \dot{\rho}(t)=-i [\cH,\rho]+\cL \rho \label{quantummasterequation}
 ~~, \eeq
where $\cL \rho$ is a superoperator describing the dissipative
dynamics. It is commonly takes the Lindblad form \cite{Lindblad}
\beq \cL \rho = \sum_i \left( -\half \{V^\dagger_i V_i,\rho   \}
+V_i \rho V^\dagger_i \right)~~,\label{LindbladEquation1}\eeq (where
$\{\cdot,\cdot\}$ are the anti-commutation relations), which ensures
positivity of the density matrix \cite{Breuer}. The $V$-operators
define the different relaxation processes induced by the
environment.

We now follow Ref.~\cite{Mukamel} and cast the density matrix into a
vector form, defining
$\vec{\rho}=(\rho_{11},\rho_{22},...,\rho_{NN},\Re \rho_{12},\Im
\rho_{12},...,\Re \rho_{1N},\Im \rho_{1N},...)$. Here the first $N$
elements account for occupation probabilities and the other elements
describe coherence between the states in the statistical mixture. It
is now a matter of rearranging the master equation into a form
$\dot{\vec{\rho}}(t)=\hat{M}\vec{\rho}(t)$, where now $\hat{M}$ is a
matrix of dimension $N^2$ which includes both the Hamiltonian and
the dissipative part of the evolution.

Due to the semi-group properties the Lindblad equation, (at least)
one of the eigenvalues of $\hat{M}$ is exactly zero \cite{Mukamel}.
The eigenvector corresponding to this eigenvalue is the steady-state
of the system, i.e. the limit $\lim_{t\to\infty}\rho(t)$. The other
eigenvalues may have an imaginary part, but all of them have a
negative real part, which correspond to the relaxation rate of the
corresponding eigenvector. Thus, by calculating the eigenvalues of
$\hat{M}$ one obtains the relaxation rates for all possible
processes. For the most general initial condition, the smallest
non-vanishing eigenvalue of $\hat{M}$ (which we call $\lambda_1$)
represents the longest relaxation rate.

\section{Application to the two-qubit system}

\subsection{Two-qubit system in a perpendicular field}
 Next we consider the
application of the above method to our model two-qubit system. The
hamiltonian, with Ising interactions, can be written as \beq \cH=-2
Js^{(1)}_z s^{(2)}_z+\textbf{B}\cdot \sum_{i=1,2} \textbf{s}^{(i)}
~~, \label{ham1}\eeq where $\textbf{s}^{(i)}$ are the two qubit
levels (i.e. spin=$\half$ particles), $J>0$ describes the
interactions and $\textbf{B}$ is the external magnetic field. This
choice of Hamiltonian is not only convenient (the Hamiltonian being
very simple), but also represents several suggestions for
realistic, spin-based quantum computers \cite{Kane,Loss}.

 For simplicity we start with magnetic field only in the $z$-direction,
i.e. $\textbf{B}=B\textbf{z}$ with $B>0$. Choosing as a basis the
four states $\ket{\uparrow\uparrow},~\ket{\uparrow\downarrow}
,\ket{\downarrow\uparrow},\ket{\downarrow\downarrow}$ (which we
number from $1$ to $4$, respectively), the Hamiltonian can be
written (up to a constant energy shift) as \beq \cH=\left(
      \begin{array}{cccc}
        -J-B & 0 & 0 & 0 \\
        0 & 0 & 0 & 0 \\
        0 & 0 & 0 & 0 \\
        0 & 0 & 0 & -J+B \\
      \end{array}
    \right)~~. \label{H} ~\eeq

In order to account for the environment, the spins are coupled to one that induces spin-flip processes, represented by the
$V$-operators in Eq.~\ref{LindbladEquation1}. Here we make two assumptions, namely (i) the spins are flipped one at a time (i.e. there is no
direct relaxation from the $\ket{\uparrow\uparrow}$ state to the $\ket{\downarrow\downarrow}$ state, etc.) and (ii) the relaxation rate between
two states is proportional to the Boltzman factor of the corresponding energy difference between the states \cite{VanKampen} (these two
assumptions on the form of the $V$-operators reflect the properties of the environment as described in the introduction). For example, the
relaxation operator from $\ket{\uparrow\uparrow}$ to $\ket{\uparrow\downarrow}$ (taking $k_B=1$) is \cite{InfoComp}  \beqa
V_{12}&=&\gamma_{12}^{1/2}\ket{\uparrow\downarrow} \bra{\uparrow\uparrow} \nonum \gamma_{12}&=& \frac{\gamma_0}{\cosh \left(\frac{J+B}{2T}
\right)} \exp\left(-\frac{J+B}{2T} \right)~~, \eeqa where $\gamma_0$ is some typical relaxation rate which represents the strength of the
qubit-environment coupling.
 For the reverse process the relaxation operator is
\beqa V_{21}&=&\gamma_{21}^{1/2}\ket{\uparrow\uparrow}
\bra{\uparrow\downarrow} \nonum \gamma_{21}&=& \frac{\gamma_0}{\cosh
\left(\frac{J+B}{2T} \right)} \exp\left(\frac{J+B}{2T} \right)~~,
\eeqa so that the condition of detailed balance is maintained, i.e.
$\gamma_{12}/\gamma_{21}=\exp \left( -\frac{\D E_{12}}{T} \right)$.
Note that one can normalize the transition rate in different ways
and still maintain detailed balance. Here we choose such a
normalization that keeps all relaxation rates finite (even at $T \to
0)$ and preserves the $B\to -B $ symmetry (i.e., does not depend on
the gauge of the Hamiltonian). However, all the qualitative results
presented in this paper equally apply for a different choice of the
normalization of the $V$-operators.

Once the form of the $V$-operators is specified, it is now a matter
of algebraic manipulation to obtain the $\hat{M}$-matrix and its
eigenvalues and eigenvectors. For the steady state we find \beqa
\rho(\infty)&=&Z^{-1}\left( \ket{\uparrow \uparrow }\bra{\uparrow
\uparrow}+ e^{\frac{-2B}{T}}\ket{\downarrow \downarrow
}\bra{\downarrow \downarrow} \right.\nonum &
&\left.+e^{\frac{-(J+B)}{T}}(\ket{\uparrow \downarrow }\bra{\uparrow
\downarrow}+\ket{\downarrow \uparrow }\bra{\downarrow
\uparrow})\right) ~,\label{SS} \eeqa with $Z$ being the partition
function of the system, which results in a pure
$\ket{\uparrow\uparrow}$ state (which is the ground state of the
Hamiltonian) in the limit of $T\to 0$.

For the above example, all the eigenvalues may be calculated
analytically. For the lowest rate we find \beq
\lambda_1=1-\frac{\sinh \left(\frac{J}{T}\right)}{\cosh
   \left(\frac{B}{T}\right)+\cosh \left(\frac{J}{T}\right)}~~,
   \label{lambda}\eeq which is $3$-fold degenerate. Two states
   contain $\rho_{14}$ and $\rho_{41}$ (i.e. a coherence between
   $\ket{\uparrow\uparrow}$ and  $\ket{\downarrow\downarrow}$), and
   the third is a mixture of all the diagonal elements
   $\rho_{11},\rho_{22},\rho_{33}$ and $\rho_{44}$. Inspection of $\lambda_1$ in the limit $T\to 0$ shows that \beq
\lim_{T\to 0} \lambda_1= \left\{ \begin{array}{c}
                                   0,~~B<J \\
                                   \gamma_0/2, ~~B=J \\
                                    \gamma_0,~~ B>J
                                 \end{array}
      \right. ~~. \label{lambda1} \eeq
This means that for $B<J$ the relaxation time from the coherent
$\rho_{14}$ state diverges, i.e. the system \emph{never} reaches its
thermal, disentangled, ground state. However, for $B\geq J$ it
relaxes to the ground state in a time scale $\tau \sim
\gamma^{-1}_0$. This can be easily explained from the following considerations. Note that the $\ket{\downarrow\downarrow}$ state can only relax
into one of the degenerate middle states, $\ket{\uparrow\downarrow}$
or $\ket{\downarrow\uparrow}$. For $0<B<J$, its energy is negative
(but higher than the ground state energy), and therefore the
relaxation rate is exponentially small. Put it differently, the
$\ket{\downarrow\downarrow}$ state is a meta-stable state which requires
an exponentially rare correlated event to escape from. For $B>J$,
however, the energy of $\ket{\downarrow\downarrow}$ is no longer negative, and hence it can relax into the ground state by a
cascade relaxation through the middle states.

\begin{figure}[t]
\vskip 0.5truecm
\includegraphics[width=8truecm]{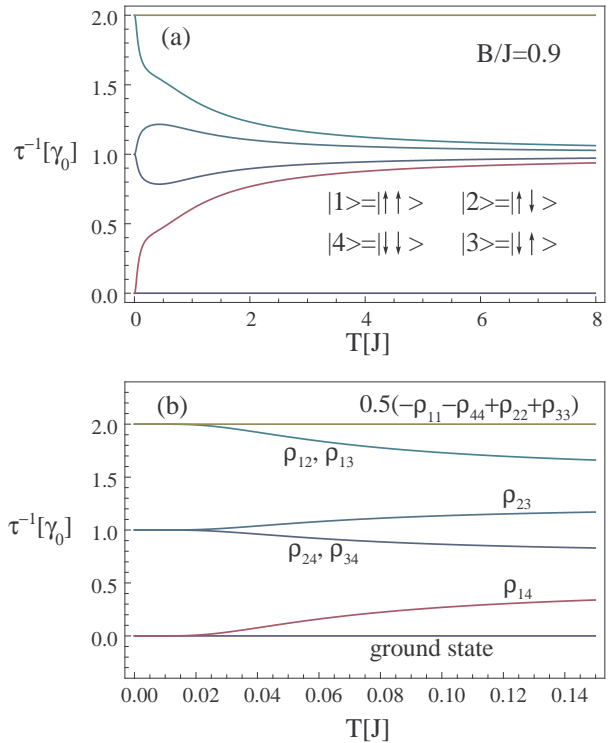}
\caption{(color online) different relaxation rates (at $B/J=0.9$) as
a function of temperature for (a) a large temperature scale, and (b)
zoomed in on the low-temperature regime. For each rate, the
corresponding density matrix element is pointed.}\label{fig1}
\end{figure}

While all the other eigenvalues are available analytically, writing
them in full form is cumbersome, and thus we present them
graphically. In Fig.~\ref{fig1} we plot the eigenvalues
(corresponding to the inverse relaxation rates for the different
states), as a function of temperature for magnetic field, $B/J=0.9$.
In Fig.~\ref{fig1}(a) we plot a wide temperature range, and we zoom
in on the low-temperature range in Fig.~\ref{fig1}(b). For each
relaxation rate the corresponding elements of the density matrix are
marked. Fig.~\ref{fig1} shows two interesting features: (i) the
relaxation rates are nonmonotonic in temperature, and (ii) they
break into groups, with only two possible time scales (
$\gamma_0^{-1}$ and $2 \gamma_0^{-1}$) at $T\to 0$.

\begin{figure}[t]
\vskip 0.5truecm
\includegraphics[width=8truecm]{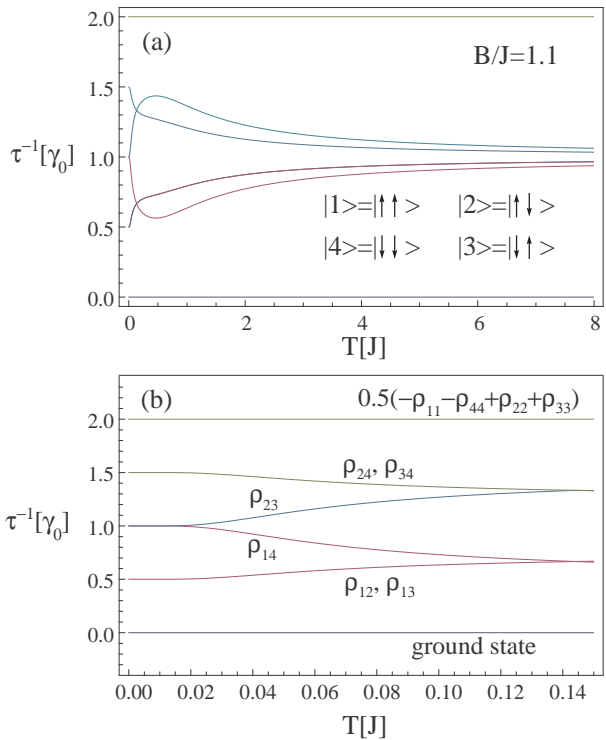}
\caption{(color online) Same as in Fig.~\ref{fig1} but for $B/J=1.1$
. }\label{fig2}
\end{figure}

For comparison, in Fig.~\ref{fig2} we plot the same for a magnetic
field $B/J=1.1$. We now find that the relaxation rates are broken
into four groups, i.e. additional time-scales appear. Interestingly,
only the $\rho_{23}$-state preserves its $T \to 0$ limit as the
$B=J$ point is crossed, while all other time-scales exhibit a
discontinuous change.

\subsection{Concurrence} In order to demonstrate the above effect
on the entanglement of the two-qubit system, we calculate the concurrence, which is a direct measure of the entanglement \cite{Hill}. For the
two-qubit system (in the basis chosen above) it is defined as $C= \max (0,\sqrt{e_1}-\sqrt{e_2}-\sqrt{e_3}-\sqrt{e_4})$, where $e_i$ are the
eigenvalues of the matrix $\hat{\zeta}$, defined by $\hat{\zeta}=\rho (s^{(1)}_y \otimes s^{(2)}_y) \rho^*(s^{(1)}_y \otimes s^{(2)}_y)$.
Clearly, the dynamics of the concurrence depend on the initial condition. For this example we choose the initial density matrix \cite{Ikram}
\beq \rho(0)=\left(
                              \begin{array}{cccc}
                                \rho_{11} & 0 & 0 & \rho_{14} \\
                                0 & \rho_{22} & 0 & 0 \\
                                0 & 0 & \rho_{33} & 0 \\
                                \rho_{41} & 0 & 0 & \rho_{44} \\
                              \end{array}
                            \right)~~, \label{rho0}\eeq
for which one can easily calculate the concurrence, $C(t)=\max(0,2(\sqrt{\rho_{14}\rho_{41}}-\sqrt{\rho_{22}\rho_{33}})$. The diagonal elements
are given by Eq.~\eqref{SS} and $\rho_{14}=\rho_{41}=1/2$. By making this choice we start with a highly entangled state ($C\approx 1$), but the
dynamics are very easy to calculate as the diagonal elements do not change at all, and the off-diagonal ones decay with the rate given by
Eq.~\eqref{lambda1}. We point out that one can start with any initial diagonal elements and obtain results similar to those presented below, since
the diagonal elements quickly relax to the steady-state (Eq.~\eqref{SS}) and do not contribute to the concurrence time-evolution.

\begin{figure}
\vskip 0.5truecm
\includegraphics[width=8truecm]{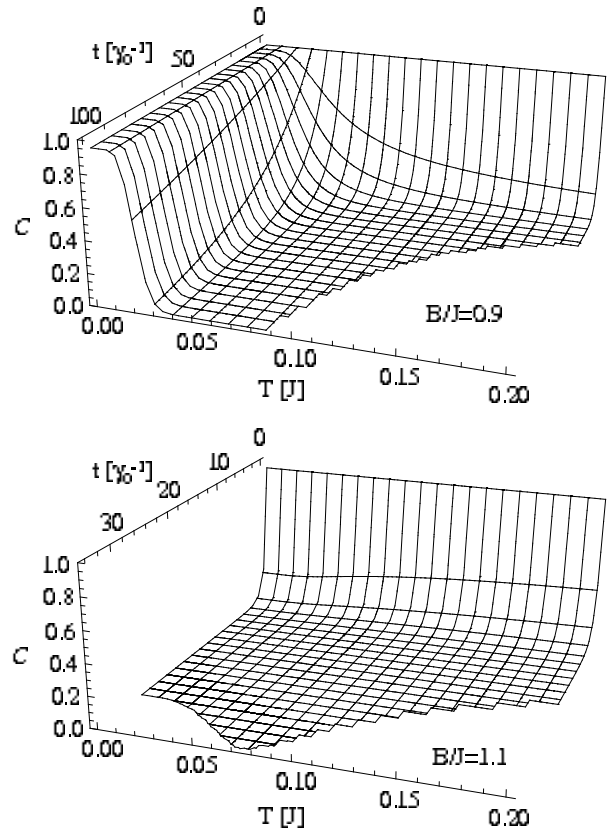}
\caption{Concurrence $C$ as a function of temperature and time for magnetic field values $B/J=0.9$ (upper panel) and $B/J=1.1$ (lower panel).
The Mesh corresponds to finite $C$, while in the blank regions $C=0$. For $B<J$ at low temperatures the concurrence practically never decays,
while for $B>J$ sudden death of entanglement appears for every temperature. }\label{fig3}
\end{figure}

In Fig.~\ref{fig3} we plot the concurrence as a function of temperature and time for magnetic field values $B/J=0.9$ (upper panel) and $B/J=1.1$
(lower panel). The mesh corresponds to finite $C$, while in the blank regions $C=0$. For $B>J$ we find that the "entanglement sudden death" is
present at all temperatures. However, for $B<J$ it becomes exponentially suppressed at lower temperatures: the concurrence practically remains
finite for all times.

In order to understand this behavior, we note that the formula for
the concurrence describes a competition between the off-diagonal
elements of the density matrix, which contribute to the
entanglement, and the diagonal elements, which "disentangle" the
state. At strictly $T=0$, the diagonal elements vanish but the
off-diagonals survive indefinitely, giving rise to an entangled
state. For very low temperature, while in the strict $t \to \infty$
limit the system becomes disentangled, this time is exponentially
long.

\subsection{Transverse field} Let us consider the effect of an
additional transverse field $\textbf{B}=B_x\textbf{x}$ (the results are identical to an additional field in the $y$-direction). A
transverse field can either represent an inherent interaction between the two qubit states, or an actual field (in the qubit
relevant Hilbert space), which is used to perform quantum operations. Since both these ingredients appear in any implementation
of a physical qubit, it is important to study their effect on the relaxation time-scales.

\begin{figure}[t]
\vskip 0.5truecm
\includegraphics[width=8truecm]{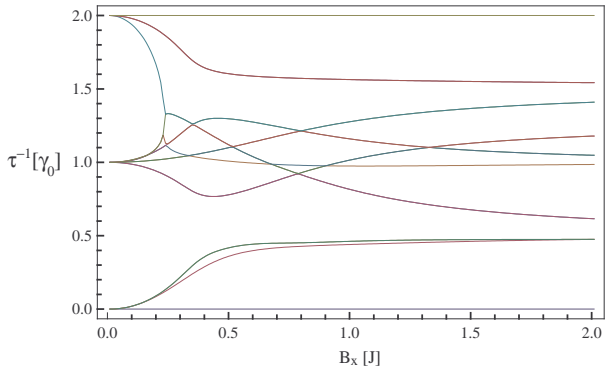}
\caption{(color online) Relaxation rates (at $T=0,B/J=0.9$) as a
function of perpendicular magnetic field $B_x$. }\label{fig4}
\end{figure}

We thus repeat the above procedure of constructing and diagonalizing the $\hat{M}$-matrix with an additional term. In
Fig.~\ref{fig4} we plot the inverse relaxation rates at $T=0, B=0.9 $ as a function of $B_x$. It is found that the infinite
relaxation time becomes finite (i.e. there is no more a meta-stable state) and that the degeneracy is lifted. This means that
applying a transverse field might give rise to entanglement "sudden-death". Interestingly, one can see that the time-scales are
not monotonic functions of $B_x$.

The conclusion arising from the above calculation is two-fold, (i) in a realistic qubit, coupling between the qubit states should
be maximally inhibited, to allow for longer entanglement life-time, and (ii) one has to take into account the fact that
performing quantum operations on the qubits will result in faster decay of entanglement.

\section{Non-Markovian effects}
In the Markovian approximation, the evolution of the system does not depend on its history. This approximation applies when the
correlation times of the thermal bath are much smaller than any timescale associated with the system under consideration.
However, there are cases where the bath reacts to the dynamics of the system over a certain memory time, $\tau_M$, and the Markov
approximation is thus not valid. In such instances, the evolution of the system would depend on its history \cite{Breuer}.
Non-Markovian effects on the entanglement dynamics have been extensively studied in recent years \cite{Dajka,Bellomo,Cao}.

In order to include non-Markovian effects, one has to include the
history of the system. In the simplest approximation \cite{Breuer},
this adds up to a form of the quantum master equation  \beq
\dot{\rho}(t)=-i[\cH,\rho]+\int^t_0 K(t-t') \cL \rho(t') \d t'
\label{nonmarkovmasterequation} ~~,\eeq  where the memory kernel
$K(t)$ defines the response of the bath to the history of the
system. Note that not every form of the memory kernel is possible,
as positivity of the density matrix may be lost \cite{Shabani}.

One can now repeat the procedure described above, and rewrite Eq.~\eqref{nonmarkovmasterequation} in a vector form, which yields the
integro-differential vector equation \beq \dot{\vec{\rho}}(t)=\int^t_0 \hat{M}(t-t') \vec{\rho}(t') \d t' ~. \label{nonmarkov2}\eeq The
time-dependence of the $\hat{M}$-matrix is such that for elements derived from the Lindblad super-operator one attaches the kernel $K(t-t')$,
and for the elements derived from the Hamiltonian one attaches a $\del$-function, $\del(t-t')$. One can now Laplace-transform
Eq.~\eqref{nonmarkov2} and obtain an algebraic vector-equation $s \rho(s)-\rho(0) = \hat{M}(s)\rho(s)$, where $s$ is the (complex) Laplace
variable. Note that the Laplace transform $\hat{M}(s)$ has a simple form, as the Hamiltonian elements are multiplied by unity (which is the
transform for the $\del$-function) and the Lindbladian elements are multiplied by the Laplace transform of $K(t)$. The formal solution of the
above equation is thus given by \beq \rho(t)=\int^{i \infty}_{-i \infty} e^{st} (s \mathbb{I}-\hat{M}(s))^{-1}\rho(0)\d s~~,
\label{nonmarkov3}\eeq where $\mathbb{I}$ is the unit matrix. From Eq.~\eqref{nonmarkov3} it can be seen that if the secular equation $\det{(s
\mathbb{I}-\hat{M}(s))}$ has solutions, the real part of these solutions defines a relaxation time-scale (via a Cauchy-like integration over
poles). One can thus obtain the relaxation times from a numerical solution of the secular equation, without the need to solve the full
non-Markovian dynamics.

In the simplest approximation \cite{Shabani,Budini,Barnett,Daffer} the memory kernel is given by an exponential form, \beq K(t)=\tau^{-1}_M
\exp(-|t|/\tau_M)~~, \eeq where $\tau_M$ is the memory-time. The Laplace transform is thus $K(s)=\frac{1}{\tau_M s+1}$, which yields a
polynomial secular equation. Note that in the limit $\tau_M \to 0$ the Markovian limit is exactly obtained. We have solved this equation
numerically, and found that it always has $N-1$ solutions with negative real part and a single solution with $s=0$, corresponding to the
steady-state. This means that for an exponential memory-kernel, one can identify different processes which have different relaxation rates. We
note that if one takes a more complicated memory kernel (say a power-law), then the secular equation becomes transcendental, with no simple
poles. In that kind of environment, one cannot simply attach different time-scales to different processes, and the full dynamics of the system
must be calculated \cite{Zwolak}.

\begin{figure}[h!]
\vskip 0.5truecm
\includegraphics[width=9truecm]{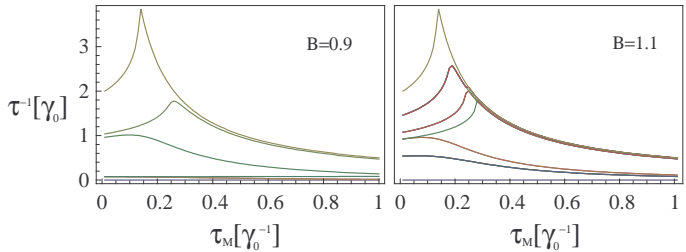}
\caption{(color online) Relaxation rates (at $T=0,B/J=0.9,0.1$) as a
function of the memory-time $\tau_M$. Note the resonant structure of
the time-scales, and the bunching of different time-scales at large
$\tau$. }\label{fig5}
\end{figure}

The above scheme can now be applied to our two-qubit model. In Fig.~\ref{fig5} the inverse time-scales are plotted as a function of the
memory-time $\tau_M$ for $B/J=0.9,1.1$. We find that the time-scales have a resonant-structure as a function of $\tau_M$, and in fact exhibit a
cusp at the resonance. In addition, at relatively large $\tau_M$, all the different time-scales merge into a single group, i.e. there is no
longer a separation of the time-scale for the relaxation processes of different coherent states.

The long memory-time behavior may be understood by the fact that in this case the bath memory dominates the relaxation processes, giving rise to
a single time-scale. The longer the memory-time, the more weight is given in the relaxation process to states far from the steady-state, and
hence the relaxation rate diminishes. However, the rise in relaxation rate at small memory-times is a surprising effect, which comes about due
to the complex nature of the interaction between the two-qubit system and the environment.

\section{Summary}
In this paper we have studied the relaxation dynamics of a model
interacting two-qubit system in the presence of an ohmic thermal
environment. The qubits were modeled as spin-$\half$ particles with
spin-spin coupling and in the presence of a magnetic field. The
environment was limited to induce only single spin-flip events.
Within this model we analytically calculated the relaxation rates of
different processes. Our main result is that disentanglement may be critically slowed down in the
$T \to 0$ limit by varying the
magnetic field, and entanglement sudden death may be completely
avoided (or at least exponentially suppressed for low temperatures).
This was explicitly shown by calculating the concurrence dynamics of
a highly entangled state for different magnetic fields. We have also
shown that a transverse magnetic field may destroy this effect.

In addition, we have introduced a simple way to include
non-Markovian effects in the calculation of the relaxation rates. We
have shown that the different relaxation rates exhibit an
interesting non-monotonic dependence on the environment memory-time,
with a cusp-like resonance. For long memory-times, we have found that the
different time-scales merge into a single time-scale.

In order to experimentally verify the results presented here, one needs an experiment with a well-controlled two-qubit system, where the
energy-levels can be controlled. A promising candidate is a system composed of two coupled two-quantum-dot qubits \cite{DiVincenzo}, where the
qubit levels may be controlled by external gates, and a high level of control of the qubit states has been already demonstrated experimentally
\cite{Petta}.

%

\acknowledgements We thank S. Saikin for fruitful discussions and Yu. V. Pershin for valuable comments on the manuscript. This
work was funded by the Department of Energy grant DE-FG02-05ER46204.

\end{document}